\input amstex

%% as per msri preprint style
\documentstyle{amsppt}
\magnification=1200
\hcorrection{.25in}
\advance\vsize-.75in

\NoBlackBoxes
\nologo

% macros

\def\diverg{\operatorname{div}}
\def\dom{\Cal D}

\def\grad{\nabla}
\def\clifh{\grad h\cdot}
\def\hess{\Cal H}
\def\Hess{\operatorname{Hess}}

\def\integers{\text{{\bf Z}}}
\def\ip#1#2{\left<#1,#2\right>}
\def\ipmu#1#2{\left<#1,#2\right>_\mu}
\def\iploc#1#2{\left(#1,#2\right)_x}

\def\lap{\triangle}
\def\real{\text{{\bf R}}}
\def\ricci{\operatorname{Ric}}
\def\volume{\operatorname{vol}}
\def\vf{\varphi}

% reference tags:  \def\name{[{\bf sym}]}
\def\Bismut{[{\bf Bi}]}
\def\Braverman{[{\bf Br}]}
\def\Bueler{[{\bf Bu}]}
\def\CaddeoMatzeu{[{\bf CM}]}
\def\CheegerGromoll{[{\bf CG}]}
\def\Chernoff{[{\bf C}]}
\def\EscobarFreire{[{\bf EF}]}
\def\Evans{[{\bf E}]}
\def\Gross{[{\bf G}]}
\def\HallOne{[{\bf H1}]}
\def\HallTwo{[{\bf H2}]}
\def\LawsonMichelsohn{[{\bf LM}]}
\def\McKeanTrubowitz{[{\bf MT}]}
\def\Petersen{[{\bf P}]}
\def\Sakurai{[{\bf S}]}
\def\Varadhan{[{\bf V}]}
\def\Witten{[{\bf W}]}

\topmatter
\title Number Operators for Riemannian Manifolds \endtitle
\author Ed Bueler  \endauthor
\date 25 August 2000 \enddate

\address Department of Mathematical Sciences, University of Alaska, Fairbanks AK 99775 \endaddress

\email ffelb\@uaf.edu \endemail

\thanks This paper was written in part at the Mathematical Sciences Research Institute in Berkeley, CA.  Research at MSRI is supported in part by NSF grant DMS--9701755. \endthanks

\abstract  The Dirac operator $d+\delta$ on the Hodge complex of a Riemannian manifold is regarded as an annihilation operator $A$.  On  a weighted space $L_\mu^2\Omega$, $[A,A^\dag]$ acts as multiplication by a positive constant on excited states if and only if the logarithm of the measure density of $d\mu$ satisfies a pair of equations.  The equations are equivalent to the existence of a harmonic distance function on $M$.  Under these conditions $N=A^\dag A$ has spectrum containing the nonnegative integers.  Nonflat, nonproduct examples are given.  The results are summarized as a quantum version of the Cheeger--Gromoll splitting theorem.
\endabstract
\endtopmatter

\document
Much of geometric analysis is the study of the ``natural second--order differential operator'' on a Riemannian manifold, the Laplace--Beltrami operator.  In quantum mechanics there is another ``natural'' second--order differential operator on $\real^n$, namely the quantum harmonic oscillator $H$.  Up to a constant scaling and shift, $H$ has whole number spectrum, thus it is a {\sl number operator}.

This paper addresses the question of whether $\real^n$ is {\sl unique} among Riemannian manifolds in this regard.  I ask: on a Riemannian manifold $(M,g)$, can one find a ``natural'' self--adjoint operator which has spectrum $\integers^+$?  Spectrum containing $\integers^+$?

If we look among Schr\"odinger operators, then the question is that of finding the potential, of course.  (See \McKeanTrubowitz~for this approach on $M=\real^1$, showing that there are in fact many such potentials.)  I approach the answer more restrictively and I look indirectly for a potential, by phrasing the question in terms of a canonical annihilation operator and a measure analogous to the ground state measure.  That is, start with a canonical $A$ acting in some (vector--valued) $L^2$ space of a specified measure.  Define $A^\dagger$ as the adjoint, and define $N= A^\dag A$.  Then ask: under what conditions is it true that $[A,A^\dagger] = 1$?

I study in this paper a particular choice of annihilation operator:
	$$A=2^{-1/2}\big(d + \delta\big)$$
acting on a Hilbert space of differential forms.  Here $d$ is the exterior derivative, and $\delta$ is its adjoint with respect to the ``usual'' $L^2$ structure on $(M,g)$.  Note $d$ is metric independent but $\delta$ depends on $g$.

Having chosen $A$, I derive sufficient conditions in the form of two partial differential equations on the logarithm of the measure density such that $[A,A^\dag]=$ constant when acting on the span of the excited states $\vf_k = (A^\dag)^k 1$.  These conditions turn out to be necessary and to be equivalent to the existence of a harmonic distance function.  In two sentences, that is the content of the current paper.

Of course, one can build a ``number operator'' $\tilde N$ on $L^2(M)$ in a thoroughly artificial manner:  for $\{\psi_k\}$ any orthonormal basis of $L^2(M)$, define $\tilde N \psi_k = k \psi_k$.  My question is rather, ``is there a number operator which has a canonical construction?''

The reader should recall the situation in $\real^1$.  The quantum harmonic oscillator with the usual choices of scale is $H=-{1\over 2} {d^2\over dx^2} + {1\over 2} x^2$ acting in the Hilbert space of functions $L^2(\real^1)$ with Lebesgue measure.  There is an equivalent operator $n+{1\over 2}$ in the weighted $L^2$ space with gauss measure
	$$\psi_0(x)^2 dx = \pi^{-1/2} \exp(-x^2) dx$$
under the unitary map (``ground state transform'') $f\mapsto f/\psi_0$.  Let
	$$a = 2^{-1/2} {d\over dx} \quad \text{ and }\quad a^\dag = 2^{-1/2} \bigg(-{d\over dx} + 2 x\bigg).$$
Then $[a,a^\dag]=1$ on $L^2(\real^1,\psi_0^2 dx)$.  Defining $n=a^\dag a$, one calculates that $n (a^\dag)^k 1 = k (a^\dag)^k 1$.  See \Sakurai~for the classic quantum harmonic oscillator.

The number operator $N = A^\dag A$ of this paper originates in the assumption that
	$$A \quad \text{ generalizes } \quad a = 2^{-1/2} {d\over dx} \quad \text{ on } \real^1.$$
In a manifold, there is only one metric--invariant first order derivative, the exterior derivative $d$.  Immediately one must deal with differential forms if one wishes to avoid making choices.  

Let $d\mu=e^{2h}\,dx$ be a measure for $h$ yet to be determined.  Let $L_\mu^2\Omega$ be the Hilbert space of square integrable forms with respect to $d\mu$.  As an annihilation operator, $\bar A = d$ itself on $L_\mu^2 \Omega$ is not a good choice!  In fact, the adjoint of $d$ on $L_\mu^2\Omega$ is $\bar A^\dag = d^* = \delta - 2i_{\grad h}$ so $[\bar A, \bar A^\dag] = d d^* - d^* d$.  This commutator is a second order operator.  There is also the self--adjoint operator $d+d^* = d+\delta - 2i_{\grad h}$ on $L_\mu^2 \Omega$.  But $\tilde A = d+d^*$ is not a good choice either since $[\tilde A, \tilde A^\dag] = 0$.

This leads me to choose $A\equiv 2^{-1/2}\left(d+\delta\right)$ which is not symmetric in $L_\mu^2\Omega$ for general $d\mu = e^{2h} dx$.  We find that $A^\dag = 2^{-1/2}\left(d+\delta + 2(dh\wedge - i_{\grad h})\right)$.  The commutator $[A,A^\dag]$ is given by the formula
	$$\left[A,A^\dag\right] = -2 \left\{ (dh\wedge - i_{\grad h}) (d+\delta) + \grad_{\grad h}\right\} + (\lap^0 h).$$
(see lemma 6 in section 2).  Thus $[A,A^\dag]$ is a first order differential operator, with zeroth--order part given by $\lap^0 h$.  Here $\lap^0$ is the nonnegative Laplace--Beltrami operator.

If we find $h$ for which $\lap^0 h$ is a constant $\alpha$ (the first partial differential equation on $h$), then we may consider $N = A^\dag A$ on the kernel of $[A,A^\dag] - \alpha$.  We call this space $\Cal P_h$.  A second partial differential equation on $h$ arises from  requiring that $A^\dag 1 \in \ker\left([A,A^\dag] - \alpha\right)$.

No further constraints on $h$ appear for $[A,A^\dag] = \alpha$ on $\Cal P_h$.  And $N$ is a number operator on the span of the excited states $\vf_k = (A^\dag)^k 1$ if the $\vf_k$ are normalizable.

In section one I state {\sl necessary and sufficient} conditions on $h$ such that the above construction yields a number operator.  See theorems 1, 2, and 3.  The conditions are the following pair of partial differential equations (1) and (2):
	$$\lap^0 h = \alpha \qquad \text{ and } \qquad \alpha h + {1\over 2} |\grad h|^2 = \gamma,$$
where $\alpha > 0$, $\gamma\in \real$ are any constants.  If $h$ satisfies (1) it follows by elliptic regularity that $h\in C^\infty(M)$.  By taking square roots (theorem 4) the above equations on $h$ are equivalent to
	$$\lap^0 r = 0 \qquad \text{ and } \qquad |\grad r| = 1$$
for a positive function $r$.  Thus a number operator with our chosen annihilation operator is equivalent to the existence of a harmonic distance function.

It then becomes clear that these number operators exist on other manifolds than Euclidean spaces.  See the examples in section 1.  In fact, in theorem 5 I interpret the existence of a number operator as a kind of ``quantum mechanical line''---a line in the geometrical sense of the Cheeger--Gromoll splitting theorem \CheegerGromoll.

Let $X$ be a constant (i.e.~translation invariant) vector field on $\real^n$, and assume $|X|=1$.  Choose coordinates on $\real^n$ so that $X={\partial\over\partial x^1}$.  Let $h_X(x) = -{\alpha \over 2} (x^1)^2$.  Then $h_X$ satisfies (1) and (2), and the associated number operator is unitarily equivalent to a quantum harmonic oscillator ``in direction $X$'' plus a free Hamiltonian (i.e.~$H_0=\lap^{(n-1)}$) in the remaining directions.  Thus there exists a solution to these equations in every dimension.  Clearly Riemannian products of euclidean space and a compact factor also allow solutions.  

The simple construction of $h_X$ on $\real^n$ relates to a literature of annihilation and creation operators on {\sl Lie groups} of compact type, which include $\real^n$.  In particular, L.~Gross (\Gross) and B.~Hall (\HallOne) initiated the study of the function space $L^2(K,\rho\,dx)$ for $K$ a Lie group of compact type and $\rho$ the {\sl heat kernel} thereon.  See \HallOne~for a definition of $\rho$ on $K$.  The measure $\rho\,dx$ is essential in creating a triple of unitary maps of Hilbert spaces on the Lie groups which correspond to the classical Hilbert spaces of the harmonic oscillator.  These maps are outside of our interest (see the survey \HallTwo) but the construction of the corresponding annihilation and creation operators, and their commutation relations, is relevant and reproduced here.

\def\axs{ a_X^* }
\def\ax{ a_X^{\phantom*} }
\def\nx{ n_X^{\phantom*} }
Let $X$ be a left--invariant vector field on $K$, so $X$ is an element of the Lie algebra of $K$.  Then
	$$\ax f = X f$$
defines an (unbounded) ``annihilation'' operator on functions $f$ on $K$.  Let $\axs = -\ax - X(\log \rho)$ be the adjoint of $\ax$ in $L^2(K,\rho\,dx)$, a ``creation'' operator.

Despite their names, $\ax$ and $\axs$ do not satisfy commutation relations analogous to $[A,A^\dag]=1$.  Rather, if $X$, $Y$ are left--invariant vector fields on $K$ then
	$$[\ax,a_Y^*] = - [X,Y] - XY(\log \rho).$$
Compare lemma 6.  Thus the spectrum of $\nx = \axs \ax$, for instance, is not easily calculated for general $K$.  If $XX(\log \rho)$ were constant on abelian $K$ (as it is on $\real^n$) then $\nx$ would be a number operator.  But a torus Lie group does not have this property.

It is clear that the replacement of the Lie group $K$ by a Riemannian manifold $M$ in this scheme is unpromising.  How to choose the global vector field $X$?  In the current paper a canonical ``isotropic'' choice $A=2^{-1/2} (d+\delta)$ replaces $a_X$.  Nevertheless a certain directionality seems to be essential, now living in the measure.  See section one.

The Lie group case motivates two questions:  I. Are there manifolds and measures for which the corresponding operators $N=A^\dag A$ are ``close to'' being number operators in the sense that (1) and/or (2) are nearly satisfied?  II. Also, are there other bundles and Dirac operators with equations corresponding to (1) and (2) which have nontrivial solutions?

Relative to the first question, consider the heat kernel $\rho=\rho(t,x,y)$ on an arbitrary complete Riemannian manifold $(M,g)$.  By definition, $\rho$ satisfies the heat equation
	$${\partial\rho\over \partial t} = -{1\over 2} \lap_x^0 \rho$$
with initial condition
	$$\lim_{t\downarrow 0} \int_M f(x) \rho(t,x,y)\,dx = f(y),$$
for all $y\in M$ and all continuous functions $f$ with compact support.  If $M$ has bounded below Ricci curvature then $\rho$ is unique and $\rho\,dx$ is a probability measure.  

The logarithm of $\rho$ satisfies an equation which converges to equation (2) in the $t\downarrow 0$ limit.  Specifically, let 
	$$h_t(x) = t \log \rho(t,x_0,x)$$
for fixed basepoint $x_0\in M$.  Then
	$$1\cdot h_t + {1\over 2} |\grad h_t|^2 = t\left({\partial h_t\over \partial t} + {1\over 2} \lap^0 h_t\right)$$
from the heat equation.  Furthermore, the famous S.~Varadhan \Varadhan~result applies to $h_t$:
	$$\lim_{t\downarrow 0} h_t(x) = -{1\over 2} d(x_0,x)^2,$$
where $d$ is the Riemannian distance in $M$.  In the small $t$, small distance limit $\lap^0 h_t \to +n$ so (1) is not approached unless $n=1$.  In fact, one can think of (1) as generically enforcing the aforementioned directionality.  The function $h_d = -{\alpha\over 2} d(x_0,x)^2$ is an a.e.~differentiable solution to (2), for any $(M,g)$ and any $\alpha>0$, $x_0\in M$.

Since $A$ is the Dirac operator for the real Hodge complex of the Riemannian manifold $(M,g)$, it is clear that other Dirac bundles (\LawsonMichelsohn) are candidates for further investigations.  Furthermore it would be appropriate to investigate vacuum states other than $1\in \Omega(M)$.

\bigskip
\heading 1. Statement of sufficient and necessary conditions \endheading

First, we give sufficient conditions on a measure $d\mu$ to construct a number operator.

\proclaim{1. Theorem} Let $(M,g)$ be a noncompact complete manifold.  Suppose $\ricci \ge -c I$ for some constant $c$.  Let $\lap^0=-\diverg \circ \grad$ be the Laplace--Beltrami operator on $M$, a nonnegative operator.  Assume that there is a real function $h \in C^2(M)$ and constants $\alpha >0$ and $\gamma\in\real$ for which the following three properties hold:
$$\align   	\lap^0 h &= \alpha, \tag1 \\
  \alpha h + {1\over 2} |\grad h|^2 &= \gamma, \qquad \text{ and } \tag2 \\
	\int_M |h|^j e^{2h} dx &< \infty \quad \text{for every integer } j \ge 0.\tag3\endalign$$

Let
	$$d\mu = e^{2h} dx,$$
a finite measure on $M$, where $dx$ is the Riemann--Lebesgue measure.  Define $L_\mu^2\Omega$ as the space of differential forms square--integrable with respect to $d\mu$.  Let
	$$A = 2^{-1/2} \left(d+ \delta\right).$$
Let $A^\dag$ be the formal adjoint of $A$ in $L_\mu^2\Omega$.  Let $\vf_k = (A^\dag)^k 1$.  Then $\vf_k \in L_\mu^2\Omega$ for all $k$.

Let $N=A^\dag A$, with domain $\Cal D_N = $ span $\{\vf_k\}_{k=0}^\infty$ (finite span).  Let $\Cal H$ be the closure of $\Cal D_N$ in $L_\mu^2\Omega$.

Then $N$ is essentially self--adjoint in $\Cal H$, $N$ has spectrum $\sigma(N)=\alpha \integers^+$, and $N\vf_k = \alpha k \vf_k$ for all $k$.  In particular, the $\{\vf_k\}$ are orthogonal.

Furthermore $A$ and $A^\dag$ satisfy the commutation relation
	$$[A,A^\dag]=\alpha\tag4$$
on $\Cal D_N$.\endproclaim

We denote the space of smooth differential forms as $\Omega(M)$, the exterior derivative as $d$, its standard (Riemann--Lebesgue measure) adjoint as $\delta$, and $\lap = (d+\delta)^2 = d\delta + \delta d$ is the Hodge Laplacian associated to $(M,g)$.  Our notation for Laplacians is then consistent: the Laplace--Beltrami operator $\lap^0$ above is $\lap$ acting on zero--forms.

In formulas,
$$\align A^\dag &= 2^{-1/2} \left(d+ \delta + 2 (dh \wedge - i_{\grad h})\right), \text{ and}\tag5\\
	  N &= {1\over 2} \lap + (dh \wedge - i_{\grad h}) (d+\delta).\tag6\endalign$$

\proclaim{(Trivial) Examples}  {\rm The hypotheses of the theorem are satisfied for $M=\real^1$ with the usual metric and $h = - {\alpha\over 2} x^2 + c$, in which case $\gamma = c \alpha$;  also for Riemannian products $M=\real^1\times F$ with $h(x,\xi)=- {\alpha\over 2} x^2 + c$, where $F$ is complete, $\ricci_F \ge -c$, and $\operatorname{vol} F < \infty$.}\endproclaim

Conditions (1) and (2) impose a certain directionality on $h$.  For instance, conditions (1) and (2) are not simultaneously satisfied by the Gaussian measure $e^{2 h_G}\,dx = e^{-|x|^2}\,dx$ on $\real^n$ for which $\lap^0 h_G = n$ but $1\cdot h_G + {1\over 2} |\grad h_G|^2 = 0$.  That is, $\alpha$ needs to be $1$ and $n$ simultaneously.

As mentioned in the introduction $h_X$, for $X$ a unit constant vector field on $\real^n$, {\sl does} satisfy (1) and (2).  But then $e^{2h_X}\,dx$ is not a finite measure as required by (3).  So we now consider $N$ constructed as above if (3) is not satisfied.  

Certainly $N$ can be defined on $\Omega_c(M)$, the smooth differential forms of compact support on $M$, by formula (6).  Then $N$ is symmetric, as can be seen by integration--by--parts in $\Omega_c(M)$.  We show that an extension $\hat N$ of $N$ is self--adjoint.  We hope to show in future work that $\sigma(\hat N) \supseteq \alpha\integers^+$.

\proclaim{2. Theorem} Let $(M,g)$ be a noncompact complete manifold.  Assume that there is a real function $h \in C^2(M)$ and constants $\alpha >0$ and $\gamma\in\real$ for which hypotheses {\rm (1)} and {\rm (2)} hold.  Define $d\mu$, $A$, $A^\dag$ and $\vf_k$ as in the previous theorem.

Let $\hat N = {1\over 2} \lap_\mu + \hess_h$, with dense domain $\Omega_c(M)$ in $L_\mu^2\Omega$.  Then $\hat N$ is essentially self--adjoint and extends $N$ in the sense that $\hat N \vf_k = N \vf_k = \alpha k \vf_k$ as smooth forms. \endproclaim

Here $\lap_\mu = \big(d + \delta - 2i_{\grad h}\big)^2 = \lap + |\grad h|^2 + A_h$ is the Hodge--Witten--Bismut Laplacian, a self--adjoint operator in $L_\mu^2 \Omega$ with a remarkable connection to the Morse theory of $M$ (\Witten, \Bismut, \Bueler).  See the appendix for the definition of the Hessian operator $\hess_h$ which acts on forms.  

Theorems 1 and 2 assume that $M$ is noncompact.  This is a necessary assumption, and one simple way to see why is to note that if $M$ is compact with $\partial M = \emptyset$ then $\int_M \lap^0 h = 0$ for any $h\in C^2(M)$.  The proof is by Green's formula.  Since $\lap^0 h = \alpha$ is the gap between eigenvalues of $N$, compactness therefore ruins $N$ as a number operator.

\proclaim{3. Theorem} Let $(M,g)$ be any Riemannian manifold, and let $h\in C^2(M)$.  Define $A = 2^{-1/2} \left(d+ \delta\right)$ and $A^\dag = 2^{-1/2} \left(d+ \delta - 2 (i_{\grad h} - dh \wedge)\right)$, as differential operators on $\Omega(M)$ (so $A$,$A^\dag$ are {\sl formal} adjoints with respect to the $L_\mu^2\Omega$ inner product).  If 
	$$\left[A,A^\dag\right] = \alpha \quad \text{(a positive constant)} \quad \text{on } \quad \Cal D = \operatorname{span}\{(A^\dag)^k 1\},$$
then $A$ and $N$ leave $\Cal D$ invariant, and furthermore
	$$\lap^0 h = \alpha \quad \text{and} \quad \alpha h + {1\over 2} |\grad h|^2 = \gamma \quad \text{for some constant } \gamma.$$\endproclaim

\demo{Proof}  (So brief we include it here\dots)  From the formula (10) for the commutator,
	$$\alpha = [A, A^\dag] 1 = -2 \big(\clifh D + \grad_{\grad h}\big) 1 + (\lap^0 h) 1 = \lap^0 h,$$
which proves (1).

Now note $A^\dag 1 = 2^{+1/2} dh$.  Since $\alpha dh = [A, A^\dag] dh = -2 \big(\clifh D + \grad_{\grad h}\big) (dh) + \alpha dh$, we see $\big(\clifh D + \grad_{\grad h}\big) (dh)=0$.  Expanding this (see (13)) gives
	$${3\over 2} \alpha dh + {1\over 2} d\big(|dh|^2\big) - {1\over 2} \alpha dh = 0,$$
that is, $d\big(\alpha h + {1\over 2} |dh|^2\big) = 0$, which is (2).\qed\enddemo

Theorem 3 shows that if we define $A$ as (proportional to) $d + \delta$, and desire that $[A,A^\dag]$ be constant, then assumptions (1) and (2) of theorem 1 are necessary.  That $[A,A^\dag]$ is constant is needed to derive in the usual way that the spectrum of $N = A^\dag A$ is (proportional to) $\integers^+$ with eigenvectors $(A^\dag)^k 1$.  Theorem 3 does {\sl not} require that $(A^\dag)^k 1 \in L_\mu^2\Omega$ or that $A$, $A^\dag$ are actual adjoints in $L_\mu^2\Omega$.

Note that (2) implies $h-{\gamma\over\alpha}\le 0$ if $\alpha>0$.

\proclaim{4. Theorem} A continuous function $h$ solves {\rm (1)} and {\rm (2)} for $\alpha>0$ iff there exists a continuous function $r$ on $M$ solving
$$\align  \lap^0 r &= 0 \qquad \text{ and}\tag7\\
 	   |\grad r| &= 1.\tag8\endalign$$
All equations are true for $x\in M$ such that $h \ne {\gamma\over\alpha}$ (respectively $r\ne 0$).  Produce $r$ from $h$ and {\sl vice versa} by $r = +\sqrt{-{2\over \alpha}\left(h-{\gamma\over\alpha}\right)}$ and $h=-{\alpha\over 2} r^2 + {\gamma\over\alpha}$.  Evidently, $h,r \in C^\infty\left(M\setminus r^{-1}(0)\right)$.\endproclaim

\demo{Proof}  (Again very brief, and by direct calculation \dots)  Recall $\lap^0 = -\diverg \grad$.  For $h$ solving (1) and (2), define $r$ as above.  Then
$$\align  \lap^0 r &= {1\over\alpha}\diverg\left({\grad h\over r}\right) = {1\over\alpha}\left(-{g(\grad r,\grad h)\over r^2} - {\lap^0 h\over r}\right) = -{1\over\alpha r} \left({-|\grad h|^2\over \alpha r^2} +\alpha \right) \\
	&= -{1\over\alpha r} \left({-2(\gamma-\alpha h)\over \alpha r^2} +\alpha \right) = -{1\over\alpha r} \left({-\alpha^2 r^2\over \alpha r^2} +\alpha \right) = 0.\endalign$$
And $\grad r = -{\grad h\over \alpha r}$ so $|\grad r|^2 = {|\grad h|^2\over \alpha^2 r^2} = 1$.

Conversely, if (8) then
	$$|\grad h|^2 = \left|\grad\left({\alpha\over 2} r^2\right)\right|^2 = \alpha^2 r^2 = 2(\gamma-\alpha h)$$
which is (2).  Adding (7), we get $\lap^0 h = \alpha \diverg(r\grad r) = \alpha \left(|\grad r|^2 - r\lap^0 r\right) = \alpha$ as well, which is (1).  Elliptic regularity for (1) and (7), respectively, implies the final statement.\qed \enddemo

Note that theorems 2 and 3 show the equivalence of existence of a number operator and an {\sl everywhere smooth} solution $h$ to (1) and (2).  Note also that $r^{-1}(0)= \{$critical set of $h\}$.

So it turns out that up to smoothness along a critical set, necessary and sufficient conditions are the existence of a {\sl distance function} ($|\grad r|=1$) which is {\sl harmonic} ($\lap^0 r=0$).  For example, on $\real^n$ the distance function $r(x)=d(x,P)$ for $P$ a codimension 1 hyperplane is harmonic.   Another way to say this is that such $M^n$ have a distance function $r$ for which $\grad r$ is incompressible.  Thus the $n-1$ dimensional submanifolds $U_s=r^{-1}(s)$, $s>0$, have constant volume in $s$.

\proclaim{(Less trivial) Examples}  Let $g(s)$, $-\infty<s<\infty$, be a smooth {\rm infinitesimally volume--preserving} deformation of complete Riemannian metrics on some $M^n$.  Assume $\ricci_{g(s)}$ uniformly bounded below.  Then $\real^1 \times M$ with metric $ds^2+g(s)$ has a number operator (in the sense of theorem 1 if $\volume g(s) < \infty$; in the sense of theorem 2 otherwise).  In particular, let $r(s,x)=|s|$.\endproclaim

A concrete example is: $\real^1\times M$ where $M=T^2$ is a torus and $g(s)$ a flat metric $f(s) dx_1^2 + f^{-1}(s) dx_2^2$ with $f$ smooth and positive.

It is known (\CaddeoMatzeu) that for $r$ a distance function {\sl from a fixed point},  nonflat $M$ with biharmonic $r$ are rare.  However there are clearly more examples with $r$ a distance function in the $|\grad r|=1$ sense.

As mentioned, $h$ satisfying (1) and (2) has a certain ``directionality''.  I assert the following geometrical analogy.

\proclaim{5. Theorem}  {\rm A.} {\rm (Cheeger--Gromoll \CheegerGromoll)}  If $(M,g)$ contains a line and $\ricci\ge 0$ then there exists a distance function $f:M\to \real$ ($|\grad f|=1$) such that $\Hess f = 0$.  Conversely, if such $f$ exist then $(M,g)$ contains a line.

\noindent {\rm B.} If $(M,g,d\mu=e^{2h}\,dx)$ has a number operator (hypotheses of theorem 3) then there exists a distance function $r:M\to\real$ ($|\grad r|=1$) such that $\operatorname{tr} \Hess r = -\lap^0 r = 0$.  Conversely, if such $r$ exists, then $(M,g)$ has a number operator (in the sense of theorems 1 and/or 2).\endproclaim

Recall that $l(t):\real\to(M,g)$ is a {\sl line} if it is a unit speed geodesic such that $d(l(t),l(s))=|t-s|$---that is, it is not just geodesic but actually distance--minimizing along its entire length.  One can easily show that if $f$ satisfies $|\grad f|=1$ and $\Hess f = 0$ then $(M,g)$ is isometric to the product $(H\times \real,g_0+dt^2)$ where $H = f^{-1}(0)$ and $g_0 = g|_H$.  See \Petersen~section 9.3.2, for instance.

Thus $(M,g)$ having a number operator in the current sense is a generalization of $(M,g)$ having a line.  The requirement $\Hess f = 0$ for distance function $f$ has been weakened to $\operatorname{tr} \Hess r=0$ for distance function $r$, and we have seen that actual splitting is not necessary.

In section 2 the formula for the commutator is proved and the space $\Cal P_h$ explored.

In section 3 we prove theorems 1 and 2.

An appendix on differential forms calculus, especially various product rules, appears at the end.

\bigskip
\heading 2. The space $\Cal P_h$. \endheading

In this section, $h\in C^2(M)$ is arbitrary. 

Let $L_\mu^2 \Omega$ be the Hilbert space of differential forms square--integrable with respect to $d\mu=e^{2h} dx$, with inner product $\ipmu\omega\nu = \int_M \iploc\omega\nu d\mu$.  Here $\iploc{\phantom{f}}{\phantom{f}}$ is the inner product in the fiber $\wedge^\bullet T_x^* M$ induced by the metric $g$, and $dx$ is the Riemann--Lebesgue measure.  Let $d$ be the exterior derivative and $\delta$ its formal adjoint {\sl computed with respect to the unweighted inner product} $\ip\omega\nu = \int_M \iploc\omega\nu dx$.  (If $M$ is oriented, we may define the Hodge $\star$ pointwise using the metric $g$, and then $\iploc\omega\nu = \star\left(\omega\wedge\star\nu\right)$.  In that case $\delta\omega = (-1)^{n(p+1)+1} \star d \star \omega$ if $\omega\in\Omega^p$.)  

Define $A = 2^{-1/2} (d+\delta)$.  Integration by parts in $L_\mu^2\Omega$ gives the formula (5) above for the formal adjoint $A^\dag$, noting $\delta(f\omega) = -i_{\grad f} \omega + f \delta\omega$.

Define $N = A^\dag A$, given by (6) above.  On functions, $N = {1\over 2} \lap^0 + (dh\wedge - i_{\grad h}) d$.  Let $\lap = (d+\delta)^2=\delta d + d\delta$, so $\lap^0 = \delta d$ is the Laplace--Beltrami operator---with sign such that $\lap^0$ is a nonnegative operator.

We compute the fundamental commutator:

\proclaim{6. Lemma}  Let $h\in C^2(M)$.  On $C^2$ differential forms,
	$$\left[A,A^\dag\right] = -2 \left\{ (dh\wedge - i_{\grad h}) (d+\delta) + \grad_{\grad h}\right\} + (\lap^0 h).\tag9$$\endproclaim

\demo{Proof}  Note $\left[A,A^\dag\right] = \left[d+\delta, dh\wedge - i_{\grad h}\right]$, and that $[d+\delta,h] = dh\wedge - i_{\grad h}$.

Then (see (A6))
$$\align 2\grad_{\grad h} \omega &= (\lap^0 h)\omega + h \lap\omega - \lap(h\omega) \\
	&= (\lap^0 h)\omega + (d+\delta) h (d+\delta)\omega - (dh\wedge - i_{\grad h})(d+\delta)\omega \\
	&\qquad - (d+\delta) h (d+\delta) \omega - (d+\delta)(dh\wedge - i_{\grad h})\omega \\
	&= (\lap^0 h)\omega - (dh\wedge - i_{\grad h})(d+\delta)\omega - (d+\delta)(dh\wedge - i_{\grad h})\omega, \endalign$$
so
	$$\left[d+\delta, dh\wedge - i_{\grad h}\right] = -2 \left\{ (dh\wedge - i_{\grad h}) (d+\delta) + \grad_{\grad h}\right\} + (\lap^0 h).\qed$$\enddemo

\proclaim{7. Definitions}
	$$D\equiv d+ \delta \qquad \text{and} \qquad \clifh \equiv dh\wedge - i_{\grad h}$$\endproclaim

In these symbols, $A = 2^{-1/2} D$, $A^\dag = 2^{-1/2} (D+2\clifh)$, and $N = {1\over 2}\lap + \clifh D$---the notation ``$\clifh$'' motivated by Clifford multiplication on forms.  Thus (9) becomes
	$$[A,A^\dag] = [D,\clifh] = -2 \left(\clifh D + \grad_{\grad h}\right) + (\lap^0 h). \tag10$$

Neither $D$ nor $\clifh$ has a degree, that is, if $\omega$ is a $p$--form, then $D\omega$ and $\clifh \omega$ are both sums of $p-1$ and $p+1$ forms.  Note, however, that
	$$D^2 = \lap \qquad \text{and} \qquad \clifh\clifh = -|dh|^2 \text{ (multiplication by the scalar $-|dh|^2$)}$$
both have degree zero.  See (A3) if needed.

Though $D$ is not symmetric for general weights $d\mu$, it is true that $\clifh$ is skew--symmetric, since $dh\wedge$ and $i_{\grad h}$ are {\sl pointwise} adjoints: $\ipmu{\clifh \omega}\nu = -\ipmu\omega{\clifh \nu}$.

\proclaim{8. Definition}
	$$\Cal P_h \equiv \ker \left(\clifh D + \grad_{\grad h}\right).$$\endproclaim
Let us be careful in what we mean by ``$\ker$'':  $\Cal P_h$ is the vector space of all $C^1$ differential forms $\omega$ for which $\left(\clifh D + \grad_{\grad h}\right) \omega = 0$.

On $\Cal P_h$ we have the commutation relation $[A, A^\dag] = \lap^0 h$, a scalar function, by construction.  On $\Cal P_h$, $N={1\over 2} \lap - \grad_{\grad h}$.

Let $\vf_k = (A^\dag)^k \vf_0$, where $\vf_0=1$.  For instance, $\vf_1 = 2^{1/2} dh$ and $\vf_2 = -2|dh|^2 + \lap^0 h.$  These $\vf_k$ become complicated for large $k$, and more importantly, $(A^\dag)^k \vf_0 \notin \Cal P_h$ for general $h$ (and $k\ge 1$).  However, the integer powers of $h$ are in $\Cal P_h$.

\proclaim{9. Lemma}  Let $h\in C^2(M)$.  Let $j\ge 0$ be an integer.  Then $h^j \in \Cal P_h$.\endproclaim

\demo{Proof}
	$$(\clifh D + \grad_{\grad h}) (h^j) = \clifh d(h^j) + \grad_{\grad h} (h^j) = j h^{j-1} \left(\clifh dh + \grad h (h)\right) = 0.\qed$$\enddemo

In fact, since $(\clifh D + \grad_{\grad h}) f = dh \wedge df$, it follows that a $C^1$ function $f$ is in $\Cal P_h$ iff $\grad h$ is parallel to $\grad f$ at every point.

\proclaim{10. Lemma}  Let $h\in C^2(M)$.  Let $j\ge 0$ be an integer.  Then
$$\align A^\dag(h^j) &= 2^{-1/2} j h^{j-1} dh + 2^{1/2} h^j dh \quad \text{and} \tag11 \\
      A^\dag(h^j dh) &= - 2^{-1/2} j h^{j-1} |dh|^2 - 2^{1/2} h^j |dh|^2 + 2^{-1/2}h^j \lap^0 h.\tag12\endalign$$\endproclaim

\demo{Proof}  Equation (11) follows from (5).  Note that $d(h^j dh) = 0$.  From (A2) and $i_{\grad f} df = |df|^2$,
$$\align  A^\dag(h^j dh) &= 2^{-1/2} \left(\delta(h^j dh) - 2 i_{\grad h}(h^j dh) \right) \\
	&= 2^{-1/2} \left(- j h^{j-1} i_{\grad h} dh + h^j \lap^0 h - 2 h^j i_{\grad h} dh\right).\qed\endalign$$\enddemo
Thus sufficient conditions for $\vf_k\in\Cal P_h$ are: (i) expression (12) for $A^\dag(h^j dh)$ can be written in terms of a linear combination of powers of $h$, and (ii) $h^j dh \in \Cal P_h$ for all $j$.

If in addition $\lap^0 h=\alpha$ is a constant, then $[A,A^\dag]=\alpha$ on $\operatorname{span}\{\vf_k\} \subset \Cal P_h$.  In fact, if $h$ satisfies (1) and (2) then $h^j dh \in \Cal P_h$ for all $j$.  Furthermore, we can reduce (12) to a linear combination of powers of $h$.  This is our strategy, implemented in the next section.

We will need this formula in the next section---use (A6) to derive it:
	$$\left(\clifh D + \grad_{\grad h}\right) (h^j dh) = {1\over 2} h^j \left(2(\lap^0 h) dh + d(|dh|^2)\right) \tag13$$
for $j\ge 0$.

\bigskip
\heading 3. Proofs of theorems 1 and 2 \endheading

\proclaim{11. Lemma}  Suppose $h\in C^\infty(M)$ satisfies {\rm (1)} and {\rm (2)} for some $\alpha >0$ and $\gamma \in \real$.  Then $(A^\dag)^k 1 \in \Cal P_h$ for all integers $k\ge 0$ and, in fact,
	$$\vf_k\equiv \left(A^\dag\right)^k 1 = \cases \sum_{i=0}^j a_{ki} \, h^i, & k=2j \text{ is even,}\\ \sum_{i=0}^j b_{ki} \, h^i dh, & k=2j+1 \text{ is odd,}\endcases\tag15$$
where $a_{ki}$, $b_{ki}$ are constants depending only on $\alpha$ and $\gamma$.\endproclaim

\demo{Proof}  Clearly $1\in \Cal P_h$.  Note $A^\dag 1 = 2^{1/2} dh$.  By (13) and (1) and (2), we get
	$$\left(\clifh D + \grad_{\grad h}\right) (dh) = {1\over 2}\left(2\alpha dh + 2d(\gamma-\alpha h)\right) = 0.$$
Thus $(A^\dag)^k 1 \in \Cal P_h$ for $k=0$ and $k=1$.

Let $k\ge 2$.  We assume as our induction hypothesis that (15) and $(A^\dag)^l 1 \in \Cal P_h$ for all $l\le k$.  We will write $a_i$ for $a_{ki}$, etc. for simplicity.

Suppose $k+1 = 2j$ is even.  Then
$$\align   (A^\dag)^{k+1} 1 &= A^\dag \left(\sum^{j-1} b_i h^i dh\right) \\
	&= \sum^{j-1} b_i (-2^{-1/2} i h^{i-1} - 2^{1/2} h^i) (|dh|^2) + b_i 2^{-1/2} h^i (\lap^0 h) \qquad \left[\text{by (12)}\right] \\
	&= \sum^{j-1} - 2^{1/2} i \gamma b_i h^{i-1} - (2^{-1/2} i \alpha - 2^{-1/2} \alpha + 2^{3/2} \gamma) b_i h^i + 2^{3/2} \alpha b_i h^{i+1},\endalign$$
by (2), which proves (15) for $k+1$ even.  Thus $(A^\dag)^{k+1} 1 \in \Cal P_h$ by lemma 9 for $k+1$ even.

Suppose $k+1 = 2j+1$ is odd.  Then
	$$(A^\dag)^{k+1} 1 = A^\dag \left(\sum^j a_i h^i\right) = \sum^j 2^{-1/2} i a_i h^{i-1} dh + 2^{1/2} a_i h^i dh$$
by (11), which proves (15).  On the other hand,
	$$\left(\clifh D + \grad_{\grad h}\right) \left((A^\dag)^{k+1} 1\right) = \sum^j b_i \left(\clifh D + \grad_{\grad h}\right) (h^i dh).$$
But by using (1) and (2) in (13), $\left(\clifh D + \grad_{\grad h}\right) (h^i dh) = 0$ as the reader should check.  Thus $(A^\dag)^{k+1} 1 \in \Cal P_h$.\qed\enddemo

\proclaim{12. Lemma}  Suppose $h\in C^\infty(M)$ satisfies {\rm (1)} and {\rm (2)} for some $\alpha > 0$ and $\gamma\in \real$.  Then
  	$$N\vf_k = \alpha k \vf_k. \tag16$$ 
If $h$ also satisfies {\rm (3)} then $\vf_k \in L_\mu^2\Omega$ for all $k\ge 0$.\endproclaim

\demo{Proof}  By lemma 11 and equation (10),
	$$N\vf_k = A^\dag(A A^\dag) \vf_{k-1} = A^\dag(A^\dag A + \alpha) \vf_{k-1} = A^\dag (N \vf_{k-1}) + \alpha \vf_k.$$
Use induction starting with $N\vf_0 = A^\dag (A 1) = 0$.  Now suppose (3).  By lemma 11, we need only prove $h^j dh \in L_\mu^2\Omega^1$ for all $j$.  But by (2)
	$$\int_M |h^j|^2 |dh|^2 e^{2h} dx = \int_M \left(2\gamma h^{2j} - 2\alpha h^{2j+1}\right) e^{2h} dx.\qed$$\enddemo

\proclaim{13. Definition}
	$$\dom \equiv \operatorname{span}_{\real} \{\vf_k\}.$$\endproclaim

This is the space of {\sl finite} linear combinations and $\dom$ is not, in general, dense in $L_\mu^2 \Omega$.

\proclaim{14. Lemma}  Suppose $M$ is complete and $\ricci \ge -c I$.  Suppose $h\in C^\infty(M)$ satisfies {\rm (1)}, {\rm (2)} and {\rm (3)}.  Then the operator $N$ is symmetric on $\dom$: if $\omega$, $\nu$ are in $\dom$ then
	$$\ipmu{N\omega}\nu = \ipmu\omega{N\nu}.$$
Thus $\{\vf_k\}_{k\ge 0}$ is an orthogonal set.\endproclaim

\proclaim{Remark}  {\rm  Before now we have had no compelling reason to believe that $N$ is symmetric on $\dom$.  Recall $N=A^\dag A$ and $A^\dag$ is the formal adjoint constructed by integration--by--parts on $\Omega_c(M)$.  But {\sl no} nonzero element of $\Cal P_h$ can be expected to live in $\Omega_c(M)$.}

{\rm First we use the fact that every first--order symmetric differential operator of the form 
	$$D_\mu = d + \delta + (\text{real zero--order terms}),$$
is essentially self--adjoint on $\Omega_c(M)$, for $M$ a {\sl complete} manifold.  Furthermore, {\sl powers} of such operators are also self--adjoint.  (See \Chernoff.  See \Bueler, section 4 for additional exposition.)}

{\rm  A B\^ochner--Lichnerowicz formula compares the growth of the Hessian operator $\Cal H_h$ to $h$ itself.  It follows that $\dom \subset \dom_{\lap_\mu}$, where $\lap_\mu$ is the square of $D_\mu$.  From formula (18) below and by the bound on $\hess_h$, we conclude symmetry.}\endproclaim

\proclaim{15. Lemma}  If $\ricci_M \ge -c I$ and if $h\in C^\infty(M)$ satisfies {\rm (1)} and {\rm (2)}, then
	$$|\Hess h|^2 \le c_1 + c_2 h \tag17$$
for $c_1=\alpha^2 + 2 c \gamma$ and $c_2 =- 2 c \alpha$.  Here $\Hess h = \grad(dh)$.\endproclaim

\demo{Proof}  For a function $u\in C^2(M)$,
	$$-\lap^0\left({1\over 2}|\grad u|^2\right) = |\Hess u|^2 - \ip{\grad u}{\grad\left(\lap^0 u\right)} + \ricci(\grad u,\grad u).$$
See e.g.~\Petersen~chapter 7 for this identity.  For $u=h$, from (1) and (2) we see
	$$|\Hess h|^2 = - \lap^0(\gamma - \alpha h) - \ricci(\grad h,\grad h) = \alpha^2 - \ricci(\grad h,\grad h).$$
Thus (17) follows from the lower bound on Ricci curvature and an additional application of equation (2).\qed\enddemo

\demo{Proof of Lemma 14}  Let $D_\mu \equiv d + \delta - 2 i_{\grad h}$, with domain $\Omega_c(M) = $ smooth forms of compact support.  It is symmetric.  Let
	$$\lap_\mu \equiv D_\mu^2 = \lap - 2 L_{\grad h}.$$
Now, \Chernoff~Theorem 2.2 implies that $D_\mu$ and $\lap_\mu$ are essentially self--adjoint on $\Omega_c(M)$, since $M$ is complete.

We relate $\lap_\mu$ to $N$.  By formula (A5),
	$$N = {1\over 2} \lap - \Xi_h D = {1\over 2} \lap - \grad_{\grad h} = {1\over 2} \lap - L_{\grad h} + \hess_h = {1\over 2} \lap_\mu + \hess_h\tag18$$
on $\Cal P_h$.  (Lemma 11 says $\dom \subset \Cal P_h$.)

Lemma 15 allows us to show $\dom \subset \dom_{\lap_\mu^*} = \dom_{\lap_\mu}$.  In fact, if $\omega = \sum^k c_i \vf_i \in \dom_N$ and $\chi \in \Omega_c(M)$, then by (18)
$$\align  \ipmu\omega{\lap_\mu \chi} &= \ipmu\omega{(2 N - 2\hess_h) \chi} \overset \star\to= \ipmu{2N\omega}\chi - 2 \ipmu{\hess_h\omega}\chi \\
	&= \sum^k i \alpha c_i \ipmu{\vf_i}\chi - 2 \ipmu{\hess_h\omega}\chi,\endalign$$
where $\star$ follows from the integration--by--parts depending on the compact support of $\chi$.

We need only show $\big|\ipmu{\hess_h \omega}{\chi}\big| \le C \|\chi\|_\mu$, for some $C>0$, to show $\omega\in \dom_{\lap_\mu^*}$.  But $\|\hess_h \omega\|_\mu \le C_n \| |\Hess h| |\omega| \|_\mu \le C_n\|\max\{1,|\Hess h|^2\} |\omega|\|_\mu \le C_n \||a + b h| |\omega|\|_\mu$ by (17), for real constants $a$ and $b$.  (The constant $C_n$ relates the operator norm of $\hess_h$ to the Hilbert--Schmidt norm of the symmetric form $\Hess h$.  See \Petersen.)  The calculation in lemma 12 shows $(a+b h) \omega \in L_\mu^2 \Omega$.

Thus if $\omega, \nu \in \dom$, then
	$$\ipmu{N\omega}\nu = {1\over 2}\ipmu{\lap_\mu \omega}\nu + \ipmu{\hess_h\omega}\nu = {1\over 2}\ipmu\omega{\lap_\mu \nu} + \ipmu\omega{\hess_h \nu} = \ipmu\omega{N\nu},$$
since $\lap_\mu$ is self--adjoint.  For the final claim,
	$$k\alpha \ipmu{\vf_k}{\vf_l} = \ipmu{N \vf_k}{\vf_l} = \ipmu{\vf_k}{N \vf_l} = l\alpha \ipmu{\vf_k}{\vf_l}$$
by lemma 12.  This implies $\ipmu{\vf_k}{\vf_l}=0$ if $k\ne l$.\qed\enddemo

\proclaim{16. Definition}  If the hypotheses of lemma 14 apply, define
	$$\Cal H \equiv \overline{\operatorname{span} \{\vf_k\}} \subseteq L_\mu^2\Omega.$$\endproclaim

Now is a good time to note:

\proclaim{17. Lemma} If $h\in C^\infty(M)$ satisfies {\rm (1)} and {\rm (2)} for $\alpha > 0$ then $\vf_k \ne 0$ for all $k\ge 0$.  In particular, $\dim \Cal H = \infty$ if $\Cal H$ is defined.\endproclaim

\demo{Proof}  We will show that $\vf_{2j} \ne 0$ for all $j$, and this suffices.  By lemma 11, $\vf_{2j} = P_j(h)$ for some polynomial $P_j$ of degree $j$.  By an easy argument in local coordinates, and using the mean value theorem, (1) implies that the range of $h$ contains a nonempty open interval.  Since the zeros of $P_j$ must be isolated, $P_j(h)=0$ identically is impossible. \qed \enddemo

\demo{Proof of Theorem 1}  Suppose $h\in C^2(M)$ satisfies (1), (2) and (3) for some $\alpha >0$ and $\gamma \in \real$.  Note that in fact $h\in C^\infty(M)$ by elliptic regularity applied to equation (1).

From lemmas 12, 14 and 17 the closure of $N$ on $\Cal H$ is unitarily equivalent to the self--adjoint multiplication operator $e_k \mapsto k \alpha e_k$ in $l^2$, under the unitary equivalence ${\vf_k \over \|\vf_k\|_\mu} \mapsto e_k$.  It follows that $N$ is self--adjoint in $\Cal H$.  The remaining claims of theorem 1 follow immediately.\qed\enddemo

\demo{Proof of Theorem 2}  By formula (18), $\hat N = N$ as differential operators on $\Cal P_h$, so $\hat N \vf_k = \alpha k \vf_k$ if (1) and (2).  (But $N$ and $\hat N$ presumably differ on $\Omega(M)$.) 

So we address the self--adjointness of $\hat N$.  There is a unitary $U:L_\mu^2\Omega \to L^2\Omega$ defined by $\omega \mapsto e^h \omega$.  Define
	$${\hat N}^U = U \hat N U^{-1} = {1\over 2} \lap_\mu^U + U \hess_h U^{-1} = {1\over 2} \lap_\mu^U + \hess_h,$$
where (theorem 4.2 of \Bueler)
	$$\lap_\mu^U = U \lap_\mu U^{-1} = \lap + |dh|^2 -\lap^0 h - 2\hess_h,$$
both acting on $L^2\Omega$.  (Since $\hess_h$ is a zeroth--order operator on $\Omega(M)$, $U\hess_h = \hess_h U$.)  By (1) and (2) ${\hat N}^U$ has the nice expression
	$${\hat N}^U = {1\over 2} \lap - \alpha h - {\alpha\over 2} + \gamma\tag19$$
as a Schr\"odinger operator with scalar potential.  Since $\gamma-\alpha h = {1\over 2}|dh|^2\ge 0$, it follows that ${\hat N}^U$ is a Schr\"odinger operator with $V=-\alpha h - {\alpha\over 2} + \gamma$ bounded below by the constant $-{\alpha\over 2}$.

We immediately get essential self--adjointness for ${\hat N}^U$ on $\Omega_c(M)$ in $L^2\Omega$ by a theorem of M.~Braverman \Braverman.  By the unitary equivalence, for $\hat N$ on $\Omega_c(M)$ in $L_\mu^2\Omega$ as well.  Braverman extends a result of I.~Oleinik to forms, and shows that (in particular) all semibounded potentials $V$ on complete Riemannian manifolds give self--adjoint $\lap+V$.\qed\enddemo

\bigskip
\heading Appendix: Differential Forms Toolbox \endheading

Though $d$ satisfies the product rule
	$$d(\omega \wedge \nu) = d\omega\wedge \nu + (-1)^p \omega \wedge d\nu,\tag{A1}$$
for $\omega$ a $p$--form and $\nu$ any other form, the formal adjoint $\delta$, a ``generalized divergence'', does not satisfy such a general product rule.  There is the special case
	$$\delta(f\omega) = -i_{\grad f} \omega + f \delta\omega \tag{A2}$$
for a function $f$.

The exterior and interior products $df\wedge$ and $i_{\grad f}$ are adjoints with respect to the usual pointwise inner product $\iploc\omega\nu = \star\big(\omega\wedge\star\nu\big)\big|_x$, and $|df|^2 = |\grad f|^2$.  The anticommutator of $df\wedge$ and $i_{\grad f}$ is scalar:
	$$i_{\grad f} df \wedge \omega +  df \wedge i_{\grad f} \omega = |df|^2 \omega. \tag{A3}$$

The next special case of a product rule for $\delta$ is:
	$$\delta(df \wedge \omega) = (\lap^0 f)\omega + \hess_f \omega - \grad_{\grad f}\omega - df\wedge \delta\omega.\tag{A4}$$
Here $\grad_{\grad f}$ is the covariant derivative, and $\hess_f$ the Hessian of $f$, both acting as derivations on forms.  In particular, $\big(\grad_X\omega\big)(Y_1,\dots,Y_n) = X(\omega(Y_1,\dots,Y_n))-\sum \omega(Y_1,\dots,\grad_X Y_i,\dots, Y_n)$ for vector fields $X,Y_1,\dots,Y_n$ and a differential form $\omega$, and
	$$\hess_f \omega\big|_x = \sum \Hess f(X_i,X_j) \phi^j\wedge i_{X_i} \omega$$
where $x\in M$ and $\{X_j\}$,$\{\phi^j\}$ are dual orthonormal bases of $T_x M$ and $T_x^* M$ respectively.  By definition $\Hess f = \grad(df)$ so $\Hess f(X,Y)=X Y f - (\grad_X Y) f$. Recall $\Hess f$ is symmetric because the connection is Riemannian.  Note $\hess_h$ is $C^\infty$ linear.

Escobar and Freire \EscobarFreire~prove (A4) and also the beautiful relation 
	$$L_{\grad f} = \hess_f + \grad_{\grad f}.\tag{A5}$$
``Cartan's formula'' says $L_X = i_X d + d i_X$ is the Lie derivative.

Finally, we use the second--order product rule
	$$\lap(f\omega) = (\lap^0 f) \omega - 2\grad_{\grad f} \omega + f\lap\omega. \tag{A6}$$

\enddocument

\bye